# EFFECT OF CRYOPRESERVATION ON THE STRUCTURAL AND FUNCTIONAL INTEGRITY OF CELL MEMBRANES OF SUGARCANE *(Saccharum sp.)* EMBRYOGENIC CALLUSES

M.E. Martínez-Montero[1*], N. Mora[1], J. Quiñones[1], M.T. González-Arnao[2], F. Engelmann[3] and J.C. Lorenzo[1]

[1]Laboratorio de Mejoramiento Genético, Centro de Bioplantas, Universidad de Ciego de Avila, CP 69450, Ciego de Avila, Cuba. * For correspondence
(Email for M.E.M-M: cubamarcose00@hotmail.com)
[2]Facultad de Biología, Universidad de La Habana, Calle 25 e/ J e I, Vedado, La Habana, Cuba.
[3]Institut de Recherche pour le Développement, 911 Avenue Agropolis, BP 64501, 34394 Montpellier Cedex 05, France.

## Abstract

In this paper, we investigated if the differences consistently noted in survival and plantlet production between cryopreserved and non-cryopreserved, control sugarcane embryogenic calluses were related to modifications induced during cryopreservation in the structural and functional integrity of cell membranes. For this, the evolution of electrolyte leakage, lipid peroxidation products and cell membrane protein contents was measured during 5 d after cryopreservation. Differences between control and frozen calluses were observed only during the first 2 (electrolyte leakage) or 3 d (lipid peroxidation products and membrane protein content) after freezing. It was not possible to link these differences with the differences noted in survival and plant production between control and cryopreserved calluses. Additional studies are thus needed to elucidate which biochemical factors, linked to survival and plantlet regeneration, are affected by cryopreservation.

## INTRODUCTION

Cryopreservation is a very useful technique for the long-term conservation of plant genetic resources and the management of large-scale production of elite genotypes by means of somatic embryogenesis (10, 29). Castillo (6) has established a propagation protocol for sugarcane by means of somatic embryogenesis, which allows mass multiplication of plants from elite varieties. However, despite its success, this protocol has an important limiting factor, which lies with the progressive loss over time of the embryogenic potential of calluses. A cryopreservation protocol based on the simplified freezing technique has thus been established for sugarcane embryogenic calluses, which allows the preservation of their embryogenic capacity (19, 20). However, it is necessary to optimize callus survival and plant regeneration after freezing to maximize the impact of cryopreservation on the overall plant production process.

In the past, cryopreservation protocols have generally been developed using an empirical approach. However, considerable advances have been made in recent years in the use of analytical tools to enhance our current knowledge of the damages induced in biological tissues by cryopreservation (9). Various biophysical, biochemical and histo-cytological techniques are available for this purpose. Such analytical tools allow the detection of those components of a cryopreservation method which cause the most damage. Usually, these studies are correlated with survival responses and viability testing (15). Once damaging events have been elucidated for each step of a protocol, it is possible to target specific measures that will reduce injury and enhance survival.

Cell membranes are one of the main targets of numerous stress events, including cryopreservation (7, 8, 15, 23). Various markers, including electrolyte efflux, lipid peroxidation products and cell membrane protein content, reflect the structural and functional integrity status of cell membranes after exposure to such stressful events (9, 15). Measurement of electrolyte leakage has been used notably for studying the desiccation and cryopreservation sensitivity of various recalcitrant seed species (24). Lipid peroxidation profiles have been used as markers of cell membrane damage during freezing of rice cell suspensions and of the coenocytic alga *Vaucheria sessilis* (5, 12). Watanabe *et al*. (27) have shown that the acquisition of tolerance to cryopreservation of rice cells was related to changes in protein metabolism. An increasing number of proteins and peptides that might contribute to freezing tolerance, by reducing the effects of dehydration associated with freezing have been identified (26). In the same way, Thierry *et al*. have observed in carrot somatic embryos the over-accumulation of boiling-stable proteins, which seems to be related to an increase in tolerance to cryopreservation (25). Besides, some enzymes, which are induced by low temperature, such as fatty acid desaturase and sucrose phosphate synthase, also contribute to freezing tolerance (14, 16). However, to our knowledge, such measurements have never been performed on sugarcane embryogenic calluses.

In this paper, we studied the effect of cryopreservation on the structural and functional integrity of cell membranes of sugarcane embryogenic calluses by measuring electrolyte leakage, lipid peroxidation products and membrane proteins. We then investigated if the differences observed for any of these parameters between control and cryopreserved samples could be related to the differences consistently noted in survival and plantlet production between frozen and non-frozen embryogenic calluses.

## MATERIALS AND METHODS

*Plant material and in vitro culture*

Embryogenic calluses were produced from sugarcane (cv. CP52-43) immature inflorescence segments (3-5 mm) according to Castillo (6). Calluses were maintained in the dark at 25 ± 2°C and subcultured every 20 d on a semi-solid Murashige and Skoog (21) medium supplemented with 30 g $L^{-1}$ sucrose, 1 mg $L^{-1}$ 2,4-dichlorophenoxyacetic acid (2,4-D), 50 mg $L^{-1}$ arginine and 500 mg $L^{-1}$ proline.

*Cryopreservation protocol and experimental design*

Calluses were subjected to the cryopreservation protocol after 120 d in culture. Cryopreservation was performed according to Martínez-Montero *et al*. (19). Calluses were transferred to 2 ml cryotubes (8 calluses/cryotube, 3-5 mm in diameter, weighing around 45-50 mg each) before freezing and subjected to a 1 h pretreatment at 0°C with a cryoprotective solution consisting of 10 % (v/v) dimethylsulfoxide and 0.5 M sucrose in standard liquid medium. Samples were frozen in a custom-made ethanol bath consisting of a polystyrene box filled with 700 ml ethanol placed in a -40°C freezer (average cooling rate: 0.5°C/min from 0 to -40°C). Ice crystallization was induced manually at -10°C in the cryoprotective medium by briefly immersing the base of the cryotubes in liquid nitrogen. Once the temperature had reached -40°C, the cryotubes were held for 2 h at this temperature before immersion in liquid nitrogen. Samples were kept for a minimum of 2 h at -196°C. Thawing took place by plunging the cryotubes in a +40°C water-bath for 1 min. Calluses were then transferred to filter paper disks and placed in Petri dishes with standard culture medium for recovery in the dark at 25 ± 2°C. Electrolyte leakage, cell membrane protein content and cell membrane lipid peroxidation-derived malondialdehyde and aldehyde contents were measured 0, 1, 2, 3, 4 and 5 d after cryopreservation on control (*i.e.* non-cryoprotected and non-frozen calluses) and frozen embryogenic calluses.

*Electrolyte leakage*

To determine the electrolyte efflux, calluses (500 mg fresh weight) were immersed in 60 mL double-distilled water, according to Sun (24). Conductivity of the imbibition water was measured after 2 h. Samples were then boiled for 10 min and cooled down to room temperature to determine total conductivity. The percentage of electrolyte leakage was calculated from the ratio: imbibition water conductivity/total conductivity. Three samples were measured for each observation date.

*Preparation of microsomal fraction*

To prepare the microsomal fraction, sugarcane calluses (2 g fresh weight) were finely ground in liquid nitrogen. Callus powder was then mixed with 20 mL extraction buffer (sucrose 0.25 M; 100 mM NaCl; 1 mM ethylenediaminetetraacetic acid; 5 mM ascorbic acid; 1 % w/v polyvinylpyrrolidone [molecular mass 24,500]; 2 mM dithiothreitol; 10 µg ml$^{-1}$ bacitracin; pH adjusted to 7.2 with 10 mM Tris). Samples were homogenized on ice with a polytron apparatus. Crude extracts were centrifuged at 10,000 g for 20 min at 4°C. The cellular debris and nuclear and mitochondria fractions were recovered in the pellet and discarded. The supernatant was re-centrifuged at 100,000 g for 1 h at 4°C. The microsomal fraction suspension was recovered in the pellet and dissolved with 150 mM KCl. One microsomal fraction preparation was made per observation date and three measurements were performed on each fraction.

*Thiobarbituric acid reactive substances assay*

To determine cell membrane lipid peroxidation-derived malondialdehyde and aldehyde contents, 0.5 mL of a trichloroacetate (20%, v/v) and thiobarbituric acid (0.5%, w/v) solution were added to 0.5 mL of the microsomal membrane fraction. Samples were briefly shaken and incubated in a water-bath at 95°C for 30 min. They were then cooled down on ice (15 min) and centrifuged at 10,000 g for 5 min at room temperature. The non-specific absorbance of the cleared product at 600 nm was subtracted from the maximum absorbance at 532 nm for malondialdehyde measurement (15), and at 455 nm for aldehydes (1). For the calculation of malondialdehyde and aldehyde concentrations, an extinction coefficient of 155 mM$^{-1}$ cm$^{-1}$ was used for malondialdehyde at 532 nm, and an extinction coefficient of 45.7 mM$^{-1}$ cm$^{-1}$ was used as an average of the extinction coefficients obtained for five aldehydes (propanal, butanal, hexanal, heptanal, and propanal-dimethylacetal) (1) and expressed in µM g$^{-1}$ fresh weight. Three samples were measured for each observation date.

*Protein assay*

Protein concentration of the microsomal membrane fraction was measured using the methods of Lowry *et al.* (18) modified by using sodium deoxycholate. Bovine serum albumin (Sigma) was used as a standard. Twenty µL sodium deoxycholate (5% w/v) was added to 120 µL microsomal membrane fraction. Subsequently, samples were supplemented with 60 µL NaOH (1 M), 800 µL reactive A (volume ratio of 1:1:98 for 1% copper sulphate: 1% sodium potassium tartrate: 2% sodium carbonate), and 100 µL Folin reagent (dilution 1:2). Absorbance was measured at 720 nm after 30 min of reaction.

*Survival and plantlet production*

Concomitantly with the above measurements, cryopreserved and non-cryopreserved calluses were allowed to further proliferate on recovery medium. Survival, which corresponded to the percentage of calluses showing size increase, was evaluated 40 d after freezing. Plant regeneration was then stimulated by transferring the calluses to medium without 2,4-D. The number of plantlets produced was recorded 80 d after cryopreservation. The evaluation of callus survival and plant regeneration involved 30 calluses each for control and cryopreserved samples.

*Statistical analysis*

The T-test, ANOVA and Duncan tests were used to analyze the results.

## RESULTS

Table 1 shows callus survival and plantlet regeneration from control and cryopreserved samples. Both survival and plantlet regeneration were significantly lower with cryopreserved calluses.

Table 1. Effect of cryopreservation (-LN: non-cryopreserved controls; +LN: cryopreserved samples) on sugarcane callus survival and plant regeneration. Data followed by the same letter are not statistically different (t-test, p<0.05).

| Treatment | Callus survival 40 d after thawing (%)* | Number of plantlets 80 d after thawing/500 mg callus fresh weight ** |
|---|---|---|
| -LN | 100 a | 270 a |
| +LN | 90 b | 150 b |

* Data were transformed for statistical analysis in accordance with $x' = 2 \arcsin(x/100)^{0.5}$.
** Data were transformed for statistical analysis in accordance with $x' = (x+0.5)^{0.5}$.

Figures 1-4 show the variations in electrolyte leakage (Fig. 1), cell membrane protein content (Fig. 2), cell membrane lipid peroxidation-derived malondialdehyde (Fig. 3) and aldehydes (Fig. 4) in control and cryopreserved calluses during the 5 d following cryopreservation.

The high electrolyte leakage observed in cryopreserved samples immediately after thawing decreased progressively over time (Fig. 1). From the third day onwards, electrolyte leakage of cryopreserved and control calluses reached similar values and remained at low and constant levels until the fifth day.

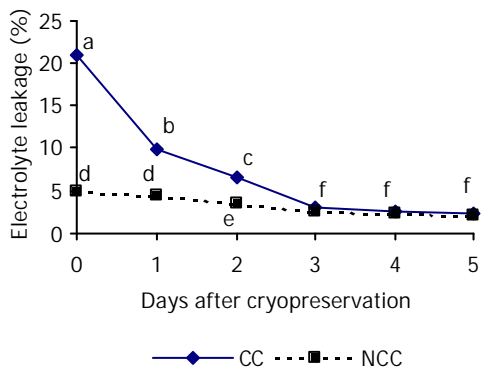
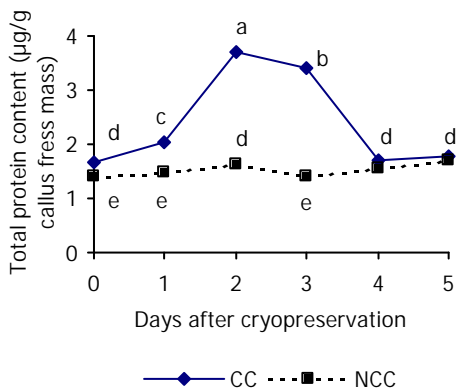

Figure 1: Changes in electrolyte leakage (% imbibition water conductivity/total conductivity) in control and cryopreserved sugarcane calluses. Data followed by the same letter are not statistically different (ANOVA, Duncan, p<0.05). CC: cryopreserved callus. NCC: non-cryopreserved callus.

Figure 2: Changes in cell membrane protein content (µg/g fresh mass) in control and cryopreserved sugarcane calluses. Data followed by the same letter are not statistically different (ANOVA, Duncan, p<0.05). CC: cryopreserved callus. NCC: non-cryopreserved callus.

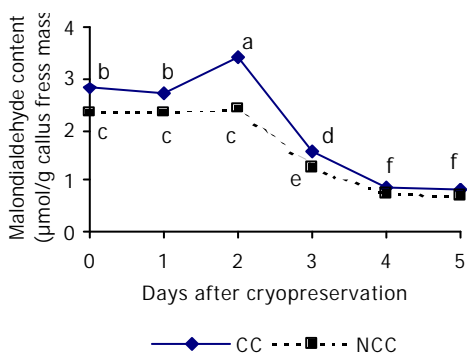
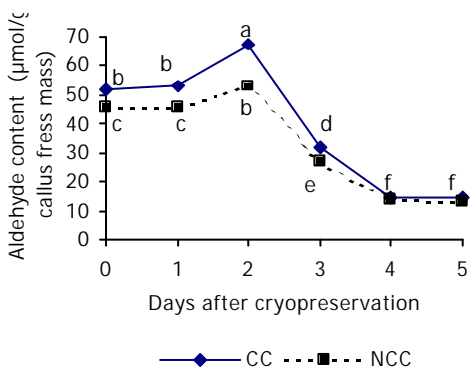

Figure 3: Changes in malondialdehyde content (µmol/g fresh mass) in control and cryopreserved sugarcane calluses. Data followed by the same letter are not statistically different (ANOVA, Duncan, p<0.05). CC: cryopreserved callus. NCC: non-cryopreserved callus.

Figure 4: Changes in aldehyde content (µmol/g fresh mass) in control and cryopreserved sugarcane calluses. Data followed by the same letter are not statistically different (ANOVA, Duncan, p<0.05). CC: cryopreserved callus. NCC: non-cryopreserved callus.

Only slight modifications were noted in the total cell membrane protein content of control calluses (Fig. 2). By contrast, a significant increase was observed in cell membrane protein content of cryopreserved calluses, with a maximum on the second day. It then decreased progressively, to reach values equivalent to those of control calluses on the fourth day.

Patterns of variation for malondialdehyde (Fig. 3) and other aldehydes (Fig. 4) were similar. Statistically significant differences between control and cryopreserved calluses were observed during the first 3 d only. Aldehyde concentration of control calluses showed no (Fig. 3) or very little (Fig. 4) variation during the first 2 d, whereas that of cryopreserved calluses reached a maximum after 2 days. After 4 days, malondialdehyde and aldehyde concentration and decreased down to 1 and 15 µmol/g, respectively, in both cryopreserved and control calluses.

## DISCUSSION

In this study, we showed that survival and plantlet production were lower with cryopreserved sugarcane embryogenic calluses in comparison with unfrozen control calluses. However, the differences observed between control and cryopreserved calluses in the parameters studied to evaluate membrane structural and functional integrity, including electrolyte leakage, total cell membrane protein content, malondialdehyde and other aldehyde content were only transitory. Indeed, they had all disappeared within 3-4 d after freezing.

Electrolyte leakage, measured to evaluate the overall effect of cryopreservation on the semi-permeability of plasma membranes, revealed a partial loss of membrane semi-permeability in callus cells. The transitory character of the electrolyte efflux observed indicates that no dramatic mechanical cell membrane injuries were caused by cryopreservation, rather only reversible lesions were induced by this treatment. As part of this dynamic process, the electrolytes released by damaged cells may have been taken up by living cells.

Freezing injury induces the production of free radicals, mainly oxygen reactive species (3). Free radicals then attack the lipid fraction of membranes, resulting in the formation of unstable lipid peroxides. These compounds breakdown to form toxic secondary oxidation products (11) such as aldehydes, including malondialdehyde and other aldehydic products. The main factors affecting sugarcane callus cell membrane damage and electrolyte efflux might thus be reactive oxygen species instead of malondialdehyde and aldehydes themselves, since the highest concentration of these compounds was reached later than the highest level of electrolyte leakage. However, it is also possible that the damage noted after cryopreservation could have been caused by the loss of cellular integrity due to the formation of ice crystals and to the cryoprotectants employed, which could damage membranes.

Our results showed that the level of malondialdehyde and other aldehydes in the microsomal fraction was higher for cryopreserved calluses than unfrozen controls, but only during the first three days after cryopreservation. Benson *et al*. (5) have obtained similar results for malondialdehyde with cryopreserved rice cell suspensions. Therefore, we suggest that freezing stress could have caused disruption and uncoupling in some metabolic pathways as reported by Fleck *et al*. (12) and Dumet and Benson (9) with other biological systems. This could have led to the production of free radicals, thus promoting lipid peroxidation in the cellular membranes of calluses at a very early post-thaw recovery stage.

Variations were also observed in control calluses, concerning mainly electrolyte leakage and lipid peroxidation. The significantly increased levels of malondialdehyde and aldehydes measured during the first 3 d in control calluses might be a result of mechanical membrane damage caused by cutting when preparing the starting material. Fleck *et al*. (12) described an increase in lipid peroxidation products after cutting algae filaments into sections. In addition, transfer of material to fresh medium itself is another stress source (28) that may cause an increase in malondialdehyde and aldehydes (4).

The concentration of lipid peroxidation products decreased from the second day onwards and reached a constant value on the fourth day in both frozen and control calluses. This decrease must have been caused by the activation of antioxidant defence mechanisms. Plants produce antioxidant molecules and have scavenging systems (ß-carotenes, tocopherol isomers, ascorbic acid, glutathione) and enzymatic free radical processing systems (superoxide dismutase, catalase, glutathione reductase, ascorbate peroxidase and various other enzymes) as a protective response to stresses (17). Those antioxidant systems are directly activated by oxidative stress and, consequently, diminish the levels of reactive oxygen species and thiobarbituric acid reactive substances in cells. Additional experiments should be performed to measure the concentration of such antioxidant molecules and the activity level of the above-cited enzymes in sugarcane embryogenic calluses in relation to cryopreservation.

An increase in cell membrane-related proteins has been described as a response to dehydration and freezing stress (2). Such proteins are produced as a protective mechanism to preserve membrane structure, ion sequestration and chaperon-like functions (25). In our experiments, the total microsomal fraction protein content was higher in cryopreserved sugarcane calluses during the first 3 d after thawing. This increase was concomitant with an increase in malondialdehyde and aldehyde concentration. We can thus offer the hypothesis that some of the proteins induced by the freeze-thaw cycle may play a role in decreasing the malondialdehyde and aldehyde levels, in addition to the other functions mentioned above.

Even though electrolyte leakage, malondialdehyde, aldehyde and cell membrane protein contents became similar in control and cryopreserved samples 4 d after cryopreservation, cryopreservation consistently rediuced callus survival and plantlet regeneration. However, lipid peroxidation products (such as malondialdehyde and aldehydes) might have impaired various cell functions in the sugarcane embryogenic calluses by cross-linking to macromolecules such as DNA and proteins to form mutagenic compounds as reported by Yang and Scaich (30) for animal cells. Moreover, the free radicals induced by freezing stress are considered both cytotoxic and genotoxic because they are capable of modifing protein structure, to form complexes with DNA and enzymes and to inhibit nucleic acid synthesis (11, 13, 22). Such impairments might have affected the totipotency of these callus cells.

In conclusion, additional studies are needed to elucidate which biochemical factors, linked to survival and plantlet regeneration, are affected by cryopreservation. Histological examination of cryopreserved calluses will also allow the identification of which cell types are preferentially damaged by cryopreservation.